\documentclass[pra,aps,twocolumn,superscriptaddress,notitlepage]{revtex4-1}
\usepackage{amssymb,amsfonts,amsmath,amstext,graphicx,hyperref}
\usepackage{tikz}
\usepackage{float}
\usepackage{graphicx}
\usepackage[normalem]{ulem}
\hypersetup{
	colorlinks=true,
	linkcolor=blue,
	filecolor=magenta,      
	urlcolor=cyan,
}
\usepackage{xcolor}

\usepackage[normalem]{ulem}
\newcommand{\cut}[1]{}

\usepackage{physics}
\def\braket#1{\langle{#1}\rangle}

\begin{document}
\title{Metrology in the Presence of Thermodynamically Consistent Measurements}

\author{Muthumanimaran Vetrivelan}
\affiliation{Department of Physics, Indian Institute of Technology-Bombay, Powai, Mumbai 400076, India}

\author{Abhisek Panda}
\affiliation{Department of Physics, Indian Institute of Technology-Bombay, Powai, Mumbai 400076, India}

\author{Sai Vinjanampathy}
\email[]{sai@phy.iitb.ac.in}
\affiliation{Department of Physics, Indian Institute of Technology Bombay, Powai, Mumbai 400076, India}
\affiliation{Centre of Excellence in Quantum Information, Computation, Science and Technology, Indian Institute of Technology Bombay, Powai, Mumbai 400076, India.}
\affiliation{Centre for Quantum Technologies, National University of Singapore, 3 Science Drive 2, Singapore 117543, Singapore.}

\date{\today}

\begin{abstract} 

Thermodynamically consistent measurements can either preserve statistics (unbiased) or preserve marginal states (non-invasive) but not both.  Here we show the existence of metrological tasks which unequally favor each of the aforementioned measurement types.  We consider two different metrology tasks, namely weak value amplification technique and repeated metrology without resetting. We observe that unbiased measurement is better than non-invasive measurement for the former and the converse is true for the latter. We provide finite temperature simulations of transmon sensors which estimate how much cooling, a resource for realistic measurements, is required to perform these metrology tasks.
\end{abstract}

\maketitle

\section{Introduction}
Quantum metrology employs non-classical resources for tasks such as  parameter estimation \cite{Giovannetti2011, beyond_noon, weak_heisenberg}, state discrimination \cite{statediscrimination, chernoff}, and hypothesis testing \cite{qchrnoff}.  The sensitivity of a metrological task achieves quantum advantage based on the non-classical resources in probe states and the choice of measurement.  Generally, such probe states are considered to be pure and measurements are considered to be ideal.  In the experiment, the probe states and measurement device are at finite temperatures.  This implies preparing pure state consumes infinite thermodynamic resources \cite{masanes2017general,wilming2017third,clivaz2019unifying}. 
As a consequence, two varieties of realistic measurements emerge namely unbiased (UB) and non-invasive (NI) measurements \cite{guryanova2020ideal}.
Incorporting these thermal resources in realistic metrology requires further study.

Hence a natural question that arises for a given metrological task, is which  type of realistic measurement is more suitable.  We answer in the affirmative by studying the utility of non-ideal measurements on different metrological schemes.  It is known that UB protects the statistics and NI protects the post-measurement state \cite{guryanova2020ideal}. Here we show a task that places a premium on statistics prefers UB measurement and the task for which post-measurement state is important prefers NI measurement. 
(section description)
At first we briefly review non-ideal measurements. We then provide the aforementioned examples and analyze why a measurement type is preferred over another. Finally, we summarize and discuss our results.


\section{Non-Ideal Measurement}

We begin with a brief review of the properties of ideal measurements and discuss why such measurements are not feasible. Consider the quantum system to be in state $\rho_{S}$ and the measuring device (pointer) $\rho_{P}$. To make an ideal measurement on the system, the eigenstates of the system observable $\{\ket{i}\}$ are correlated to the orthogonal states of a pointer $\{\ket{\psi^i_n}\}_n$. This is done by jointly evolving the system and pointer from state $\rho_S\otimes\rho_P\rightarrow \rho_{SP}$. Following this a projective measurement is performed on the pointer using $\Pi_i=\sum_n \ketbra{\psi^i_n}$ and the system state is inferred. If the correlation is perfect and the pointer is observed in  $\ket{\psi^i_n}$, we conclude the system to be in state $\ket{i}$. Such ideal measurements have three fundamental properties namely \textit{unbiasedness, non-invasiveness} and \textit{faithfulness} \cite{guryanova2020ideal}. A measurement is said to be non-ideal if any one of the properties is not satisfied. 

The \textit{unbiased} property states that the pre-measurement statistics of the system are accurately reflected by the post-measurement pointer statistics , i.e.,
\begin{equation} 
\Tr[\mathbb{I} \otimes \Pi_{i} \rho_{SP}] = \Tr[\ketbra{i} \rho_{S}] , \ \ \forall \ i.
\label{unbiased}
\end{equation}
The second property desired for ideal measurements is \textit{non-invasiveness}. This property says that the measurement interaction should not change the measurement statistics for the system namely,
\begin{equation} 
\Tr[\ketbra{i}\otimes \mathbb{I} \rho_{SP}] = \Tr[\ketbra{i} \rho_{S}], \ \ \forall \ i. 
\label{non-invasive}
\end{equation}
Finally, a measurement is \textit{faithful} if there is a one-to-one correspondence between the pointer outcome and the post-measurement system state namely,
\begin{equation} 
C(\rho_{SP})=\sum_{i} \Tr[\ketbra{i} \otimes \Pi_{i} \rho_{SP}] = 1, \ \  \forall \ i.  
\label{faithful} 
\end{equation}
This property suggests there should be perfect correlation between system eigenstates ($\{\ket{i}\}$) and pointer states ($\{\ket{\psi^i_n}\}_n$). 
Such measurements are possible only if  $\mathrm{rank(\rho_P)} \leq \mathrm{dim(\rho_P) /dim( \rho_S)}$ \cite{guryanova2020ideal}, i.e., an ideal measurement requires a non-full rank pointer state.  As the laws of thermodynamics forbid the preparation of a non-full rank state with finite resources, a realistic measurement is always non-ideal. 
\begin{figure}[]
    \centering
    \includegraphics[width=\linewidth]{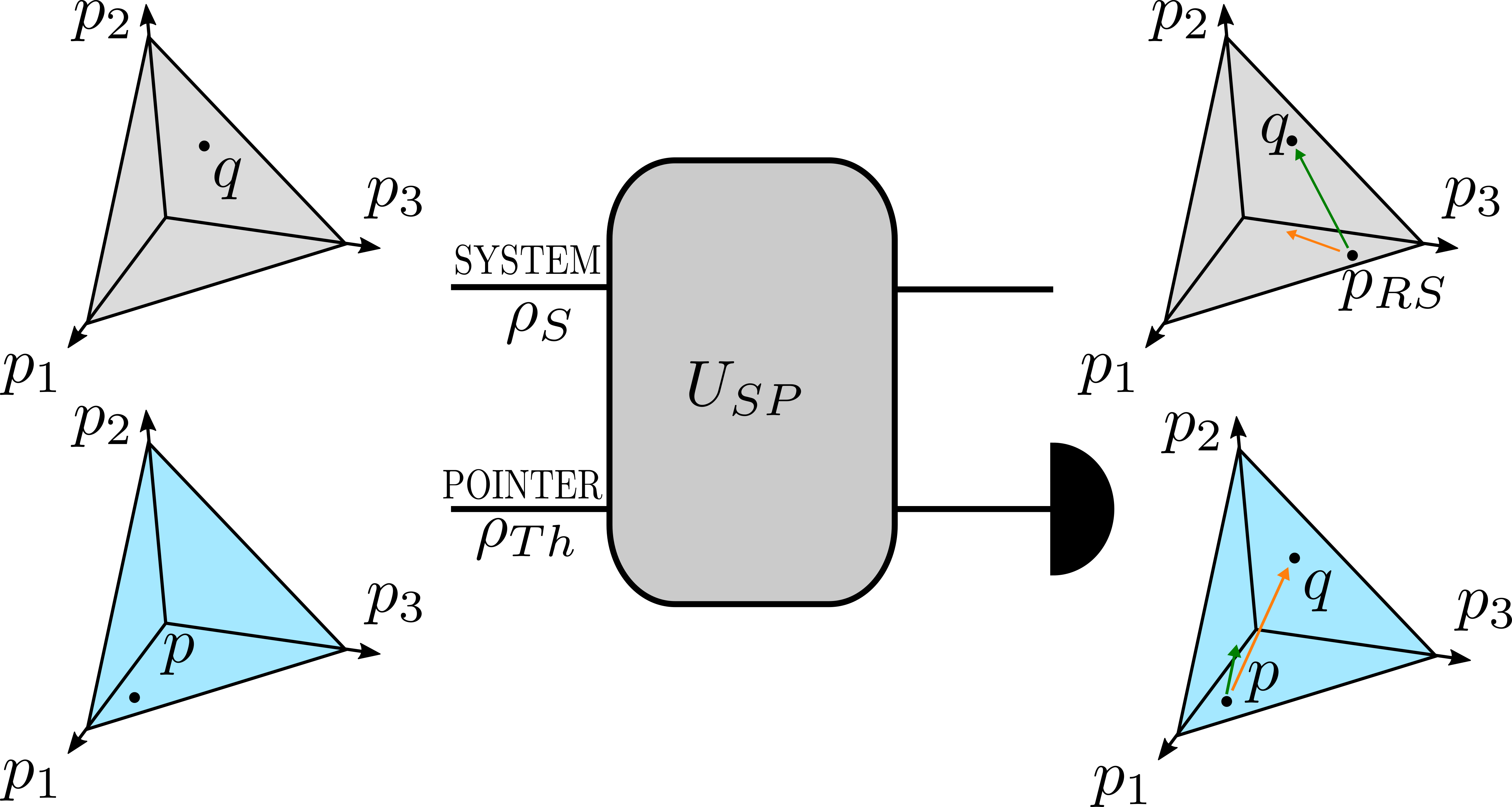}
    \caption{An illustration of the change in the probabilities of the system and pointer states in a fixed basis across a measurement as described in the text. The probability vector associated with the system statistics (grey) and for that of the pointer (blue) are shown across a correlating unitary followed by a measurement. In the space of the system statistics, the statistics ($p_{RS}$) of the reduced state is the same as the system statistics ($q$) for NI measurements (green dashed line) and points to a different probability vector for UB measurements (orange solid line).  In the case of the pointer, the pointer statistics ($p$) correctly points $q$ for UB but not for NI.}
    \label{niub}
\end{figure}
At finite temperature, the system and the pointer evolve jointly to get maximize faithfulness but the correlation will not be perfect ($C(\rho)<1$). A measurement cannot be unbiased and non-invasive as both properties together would also imply faithfulness \cite{guryanova2020ideal,debarba2019work}. Hence a maximally faithful measurement can either be unbiased (UB) or non-invasive (NI). 
The UB measurement replicates the system statistics through pointer but changes the statistics of reduced system after measurement. Due to this, the system needs to be discarded or reset to measure again. On the other hand, the NI measurement preserves the statistics of the system in the reduced system state post-measurement at the expense of the pointer statistics being different from the system statistics is shown in  Fig.~(\ref{niub}). These properties associated with different non-ideal measurements make them favorable in different scenarios. In the next section, we begin by giving the effect of non-ideal measurements in weak value amplification (WVA), where we show that UB is favored over NI measurement scheme. 


 \section{Weak Value Amplification Favors Unbiased Measurements}
WVA is a metrological scheme that uses post-selection to amplify small signals \cite{vaidman2017weak}. It involves preparing quantum states and post-selecting the evolved state in a specific final state. It has a well-known trade-off between post-selection probability and amplified weak value \cite{pangbrun}.  Though rejected measurements leave out some quantum information \cite{knee2014}, WVA has demonstrated better metrological performance in the presence of certain noises \cite{howell}.  WVA has proven useful in disparate experiments such as amplifying optical non-linearities \cite{feizpour}, measuring ultrasmall time delays of light \cite{bruder},  detector saturation and measuring small frequencies \cite{dixon}, maintaining its relevance in modern quantum metrology. From a thermodynamic perspective, the reliance of WVA on pure states precludes us from implementing it in finite dimensional physical systems as explained below.  We hence investigate WVA with constraints in preparation of states and realistic non-ideal measurements. For this reason, we summarize WVA in a way that readily generalizes to mixed states. 

The parameter $g$ is coupled through system $A$ and an ancilla degree of freedom known as a \textit{meter} $B$ via the Hamiltonian $H =g{A}\otimes {B}$. We note that this meter degree of freedom accumulates the effect of postselection and is separate from the pointer introduced to model thermodynamic resource costs of performing measurements. The system and meter are prepared in an initial state $\rho_{s}\otimes\rho_{m}$ and are evolved for a small time $\tau$, where $g\tau \ll 1$ under the influence of the Hamiltonian above.  The system is post-selected by a nearly orthogonal pure state $\ket{\psi_{f}}$ (generalized below to mixed states), such that the average of the expected spin can be larger than the mean. 

\begin{equation}
\mathcal{A}_{\mathrm{w}} = \frac{\bra{\psi_{f}}A\rho_{s}\ket{\psi_{f}}}{\bra{\psi_{f}}\rho_{s}\ket{\psi_{f}}}=\frac{\bra{\psi_{f}}A\rho_{s}\ket{\psi_{f}}}{P_s}. \label{ref4}
\end{equation}
This effectively evolves the meter state as $\exp({-ig\tau\mathcal{A}_{w}{B}})\rho_m\exp({ig\tau\mathcal{A}_w{B}})$.  The postselection of the system state in $\ket{\psi_{f}}$ depends on a projective measurement onto a pure state, which as noted above is thermodynamically inconsistent.

To include the effect of non-ideal measurements, we introduce a pointer degree of freedom.  Our setup consists of three parts i.e. system($\rho_s$) of dimension $d_s$, meter($\rho_m$) with dimension $d_m$ and pointer($\rho_p$) with dimension $d_p$. The system and pointer are governed by local Hamiltonian $H_s$ and $H_p$ respectively, where
\begin{equation}
    \begin{aligned}
    H_S&=E_s\ketbra{\psi^{\perp}_i}, \hspace{5mm}
    H_P&=E_p\ketbra{\psi^{\perp}_f}.\\
    \end{aligned} \label{wva1}
\end{equation}

To post-select the system through a pointer, the dimension of pointer should be an integer multiple of the system. The system is weakly correlated to the meter and strongly correlated to the pointer. The strong projective measurement and post-selection by pointer gives the amplified reading in the meter.  To simulate non-ideal measurement we prepare initial system and pointer to be in a thermal state, 
\begin{equation}
\begin{aligned}
    \rho_{S} &=  q \ket{\psi_{i}}\bra{\psi_{i}} + \Bar{q} \ket{\psi_{i}^{\perp}}\bra{\psi_{i}^{\perp}}, \\\rho_P&=p\ket{\psi_{f}}\bra{\psi_{f}} + \Bar{p} \ket{\psi_{f}^{\perp}}\bra{\psi_{f}^{\perp}},
\end{aligned} 
\end{equation}
where $\Bar{q}/q=e^{-\beta H_S}$, $\Bar{p}/p=e^{-\beta H_P}$ and $\beta=$1/T.   Prior to performing a measurement, we correlate system and pointer in measurement basis using a suitable unitary matrix.  To perform a suitable measurement NI (or UB), we use $U_{corr}=I_m\otimes\sum_{j}|j\rangle \langle j|\otimes \Tilde{U}^{j}$ (or $U_{corr}=I_m\otimes\sum_{ij}|i\rangle \langle j|\otimes \left|j\right\rangle\left\langle i\right|\Tilde{U}^{j}$ ). The unitary $\Tilde{U_j}$ is adjusted such that the measurement is maximally faithful. The non-ideal measurement is then simulated by doing projective measurement on pointer which is an instance of our \textit{Heisenberg's cut} \cite{haroche2013nobel}.
Given that the pointer is not pure, the desired near-orthogonal prjection of the system state happens probabilistically, the other outcome being less favorable and governed by the non-zero temperature pointer. This causes post-selected meter state to be a convex mixture of two outcomes, one which we term the \textit{true positive} (kicked state) and another which we label \textit{false positive state}. This is given by
\begin{equation}
    \rho_m^{PS(NI)}=(pqP_s+\Bar{p}\Bar{q}P_s)\eta_1+(p\Bar{q}\Bar{P_s}+\Bar{p}q\Bar{P_s})\eta_2.
    \label{psni}
\end{equation}
Here $\eta_1=\exp({ig\mathcal{A}_w\hat{B}})\rho_p \exp({-ig\mathcal{A}_w\hat{B}})$ is the weak value amplified meter state and $\eta_2=\exp({ig\mathcal{A}^{\perp}_w\hat{B}})\rho_p \exp({-ig\mathcal{A}^{\perp}_w\hat{B}})$ is non amplified meter state where
\begin{equation}
\mathcal{A}^{\perp}_{\mathrm{w}} = \frac{\bra{\psi^\bot_{f}}A\rho_{s}\ket{\psi^\bot_{f}}}{\bra{\psi^\bot_{f}}\rho_{s}\ket{\psi^\bot_{f}}}.
\end{equation}
The detailed calculation of kicked state is given in Appendix A.  Likewise, for unbiased measurement, the post-selected meter state will be 
\begin{equation}
\rho_m^{PS(UB)}=qP_s\eta_1+\Bar{q}\Bar{P_s}\eta_2.
    \label{psunb}
\end{equation}
As $\mathcal{A}^{\perp}_w\ll \mathcal{A}_{w}$, we can approximate $\eta_2\approx\rho_m$. The post-selected states in Eq.(\ref{psni}) and Eq.(\ref{psunb}) will have reduced amplification compared to the ideal scenario. 

From Eq.~(\ref{psunb}), we can note that the trade-off between post-selection probability and amplified weak value implied in Eq.~(\ref{ref4}) still persists.  This trade-off further depreciates in Eq.~(\ref{psunb}) due to preparing the initial state in the mixed state and performing the non-ideal measurements for post-selection.  Besides this, another variety of tradeoff is represented by the fact that post-selection implies that some measurements outcomes are ignored.  Measuring the system onto a rarely post-selected state and ignoring other possible final states leads to discarding potential information available from the system. This leads to sub-optimal information theoretic performance of the WVA scheme compared to the conventional scheme in which the information from all possible final states are collected. To measure performance in the ideal scenario, the quantum Fisher information for the post-selected scheme $\mathcal{I}_{PS}(g)$ and the total quantum Fisher information $\mathcal{I}(g)$ due to the initial state are compared. Hence the quantum Fisher information obtained by the WVA scheme will always be less than $\mathcal{I}(g)$, and the ratio $\mathcal{I}_{PS}(g)/\mathcal{I}(g)$ is always less than unity \cite{pangbrun}.  In a similar spirit,  we can compare the ratio of the quantum Fisher information in presence of thermal resources $\mathcal{I}_{TH}(g)$ to the ideal post-selected $\mathcal{I}_{PS}(g)$.

 The quantum Fisher information is evaluated from the Bures distance between $\rho_{g}$ and $\rho_{g+dg}$ \cite{braunstien} defined as 
$D_{B}^2(\rho _{g},\rho _{g+dg})=2(1-{\sqrt  {F(\rho _{g},\rho _{g+dg})}}),$
where $\rho_{g}$ is the state containing the information of parameter $g$ and $F(\rho _{g},\rho _{g+dg})$ is the fidelity defined as ${\displaystyle F(\rho _{1},\rho _{2})=\left[{\mbox{tr}}({\sqrt {{\sqrt {\rho _{1}}}\rho _{2}{\sqrt {\rho _{1}}}}})\right]^{2}.}$ The  quantum Fisher information as a measure of sensitivity is related to the curvature of the Bures distance at the parameteric value $g$ by the formula $\mathcal{I}(g) = - \partial_{g}^2 D_{B}^2(\rho _{g},\rho _{g+dg}).$ For the ideal WVA process, we assume $\braket{B} = 0$ and the $\mathcal{I}_{PS}(g)$ \cite{davidovich} is given by

\begin{equation}
    \mathcal{I}_{PS}(g) = 4 P_{s}\vert{\mathcal{A}_{\mathrm{w}}}\vert^2 (1-\vert g \mathcal{A}_{\mathrm{w}}\vert^2 Var(B)), \label{qf1}
\end{equation}
where $Var(B)$ is with respect to initial meter state.  The post-selected Fisher information $\mathcal{I}_{PS}(g)$ is always less than unity due to the rarity of the desired measurements. Now for non-ideal measurements, the $\mathcal{I}_{TH}(g)$ is calculated as (see appendix A)
\begin{equation}
    \mathcal{I}_{TH}(g) \approx 4 P_{M} \vert{\mathcal{A}'_{\mathrm{w}}}\vert^2 (1-\vert g \mathcal{A}'_{\mathrm{w}}\vert^2 Var(B)).
\end{equation}
Here $\mathcal{A}'_\mathrm{w}$ is true weak value  amplification defined as 
\begin{equation}
\mathcal{A}'_{\mathrm{w}}  =  \frac{\mathcal{A}_{\mathrm{w}}}{1+\delta_{M}}, \label{ref12}
\end{equation}
\begin{figure}[]
    \centering
    \includegraphics[width=\linewidth]{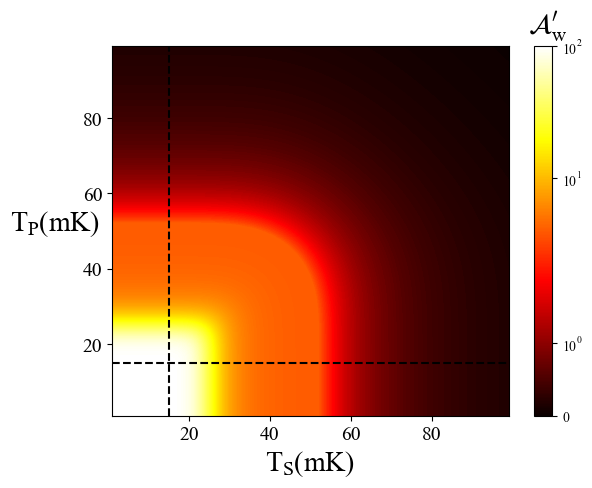}
    \caption{Contour plot of true amplification for system and pointer qubits composed of transmons at $E_p=E_s=5$GHz for NI measurements.  The amplification is denoted by the colorbar and can be seen to depend on both initial state and pointer temperatures plotted along the two axis. The dotted line represents the lowest temperature achievable by the dilution refrigerator. The achievable amplification is hence the intersection of the region to the right of the vertical and above the horizontal dotted lines.}
    \label{fig:1}
\end{figure}

\begin{figure}[]
    \centering    \includegraphics[width=\linewidth]{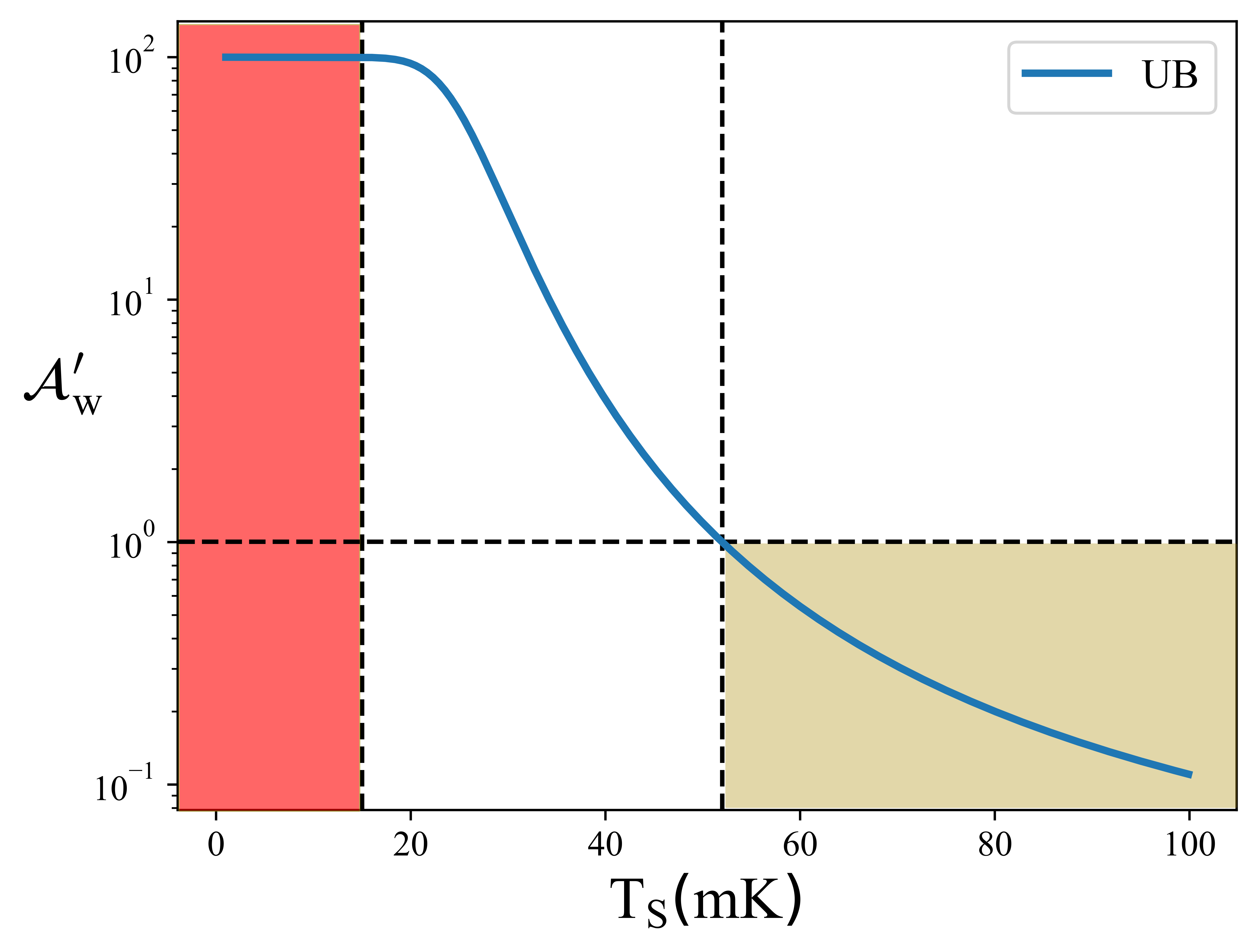}
    \caption{True amplification plotted as a function of the system temperature for UB measurement with $E_p=E_s=5$GHz. The amplification depends only on the initial state temperature in UB measurement as discussed in the text. The true amplification $\mathcal{A}'_\mathrm{w}<1$ for $T_{S} > 52\mathrm{mK}$, which is shaded in brown and denotes deamplification due to the effect of temperature. The red region shows the temperature range unattainable in the experiment considered in the text.}
    \label{fig:2}
\end{figure}
 
\begin{equation}
\text{where }\delta_{M} = \begin{cases} 
\frac{\Bar{q}\Bar{P}_{s}}{qP_{s}} &  \text{for unbiased,} \\
\frac{\Bar{p}q\Bar{P}_{s}+p\Bar{P}_{s}\Bar{q}}{pqP_{s}+\Bar{p}\Bar{q}P_{s}} \label{ref13} & \text{for noninvasive}.
       \end{cases}
\end{equation}
As $\delta_M>0$ at any non-zero temperature, the true amplification is always less than the ideal amplification. The post-selection probabilities ($P_{M}$) also depend on the system and pointer as

\begin{equation}
P_{M} = \begin{cases} 
qP_{s} &  \text{for unbiased,} \\
pqP_{s}+\Bar{p}\Bar{q}P_{s} & \text{for noninvasive}.
\end{cases}
\label{eq14}
\end{equation}

In summary, WVA is in general affected by rare post-selection given in Eq.~(\ref{ref4}).  In addition, the initial mixed state and the choice of non-ideal measurements performed at the post-selection further decrease the amplification given in Eq.~(\ref{ref12}).  From  Eq~.(\ref{ref12}) and Eq~.(\ref{ref13}), we observe that the amplification due to UB measurement depends only on the system temperature, and for NI measurements it depends on both system and pointer temperatures. To compare these measurement schemes in a concrete setting, we consider the original thought experiment \cite{spin_100} by Aharanov, Albert and Vaidman where the result of a measurement of a component of a spin$1/2$ particle was amplified by 100, now implemented with thermal resources.  Furthermore we situate this thought experiment in the IBM transmon qubit and investigate the amplification obtained for different non-ideal post-measurements. 

We initialize the system in a thermal state and perform the non-ideal measurement to estimate the steady state temperature to which the IBM transmon qubit has to be cooled for getting desired amplification.  The IBM transmon qubit works at the characteristic frequency around $5-5.4 \mathrm{GHz}$ \cite{chow} and also consists of a dilution refrigerator that can cool down the system to $15\mathrm{mK}$ \cite{dilution_refrigerator}. We note that while we are taking $15\mathrm{mK}$ as the lower limit for the current discussion, the real lower limit on the temperature to which a given system can be cooled depends on hardware constraints which should be appropriately considered while applying our analysis to experiments. The original thought experiment \cite{spin_100} discussed consists of a system prepared in a pure state $\rho_{i} = \ket{\downarrow}\bra{\downarrow}$. The system is then evolved by the Hamiltonian $A = \sigma_{x}$ and finally post-selected onto the state $\ket{\psi_{f}} = \cos(\theta)\ket{\uparrow}+\sin(\theta)\ket{\downarrow}$, gives the amplification $\mathcal{A}_{\mathrm{w}} = cot(\theta)$.  Choosing $\theta = 0.01$, the parameter $g$ is amplified  $\mathcal{A}_{\mathrm{w}} \approx 100$ times.  Performing this experiment demands stringent cooling requirements.  To demonstrate this, we simulated the non-ideal amplification obtained by WVA scheme performed on the IBM transmon setup as a function of the temperatures of system and pointer. Fig.~(\ref{fig:1}) shows the simulation for NI measurement and Fig.~(\ref{fig:2}) shows the simulation for UB measurement.  The amplification obtained from NI measurement is affected by both system and pointer temperatures as shown in Eq.(\ref{ref13}). The pointer states prepared above $30\mathrm{mK}$ prove to be detrimental for weak value amplification ($\mathcal{A}'_\mathrm{w} < 1$) even for system state prepared at very low temperature. For UB measurement, the initial system state prepared below $52\mathrm{mK}$, shows an amplification ($\mathcal{A}'_\mathrm{w} > 1$) and states prepared below $20\mathrm{mK}$ attain near desired amplification.  This is irrespective of the temperature in which the pointer states are prepared as shown in Eq.(\ref{ref13}).  As cooling costs energy \cite{clivaz2019unifying,allahverdyan2011thermodynamic,vinjanampathy2016quantum}, the resource cost for UB measurements is much more favorable than NI measurements.

In summary, we observe that UB measurements were favored over NI due to the resource requirements associated with cooling quantum systems. This is the consequence of properties of UB measurements where pointer replicates system statistics irrespective of its own temperature. From Eq.(\ref{ref13}) and Eq.(\ref{eq14}) it can be observed that true amplification and post-selection probability do not depend on pointer statistics, as denoted also in Fig~(\ref{niub}). Hence the UB measurements are preferred if getting the correct statistics of the system is more important. In contrast to this, there are metrology protocols which prioritize reduced states over measurement statistics. In the next section, we considered one such example and show how  NI measurements have an operational advantage over UB measurement.

\section{Sequential Metrology Favors Non-Invasive Measurements}

  While conveyer belt models of metrology \cite{conveyor} imply that the initial state is an infinite free resource, sequential metrology schemes that prevent resetting account for the role played by the initial coherences of the state. The sequential metrology scheme without resetting involves initializing the probe state once followed by cycles of evolution and measurement of the final state. This is different from repeated interaction schemes where an initial probe state interacts with the parameter repeatedly before being measured once \cite{seq_wiseman}. Recently a sequential metrological scheme without resetting has been proposed to beat the shot noise limit without exploiting entanglement resources in probe states \cite{montenegro2022sequential}. A similar scheme was used in estimating the temperature of the thermal reservoir by sequentially measuring the probe states in contact with the reservoir without resetting \cite{burgarth}. In such settings, the statistics need to be preserved for the system and conveyed through a pointer before the next measurement. Simultaneously, the post measurement state of the system needs to be able to continue to acquire information about the unknown parameter, which is affected by the type of realistic measurement used. Hence a natural question that arises is whether sequential metrology without resetting places a natural preference on statistics or reduced states. We now compare sequential metrology with different thermodynamically consistent measurements discussed before.
  
Once again, we consider the transmon qubit initially in the thermal state with transition frequency $5\mathrm{GHz}$ at temperature $\mathrm{T_{S}} = 100 \mathrm{mK}$. The system is evolved according to Hamiltonian $H = \theta \sigma_{x}$, acquiring the unknown phase $\theta$ whose inference proceeds by a measurement by correlating to a pointer. As outlined above, the sequential nature of this metrology scheme is the repeated interaction and measurement. We start with system initialized in state $\rho_{S,0}$ which will be evolved and measured repeatedly with measurement operators $M_{1,2} = [\ket{\downarrow}\bra{\downarrow}, \ket{\uparrow}\bra{\uparrow}]$ acting on the pointer state. Each sequence consists of a pair of unitary evolution which gains phase, followed by a measurement event. There are $N_{s}$ number of sequential measurements performed on each initialized system and this whole sequence of measurements is repeated $\nu$ times for statistical inference. For instance, $\rho_{S,i}$ represents the initial system state at the $i^{th}$ step of the sequence of evolution and measurements.

\begin{figure}[]
    \centering
    \includegraphics[width=\linewidth]{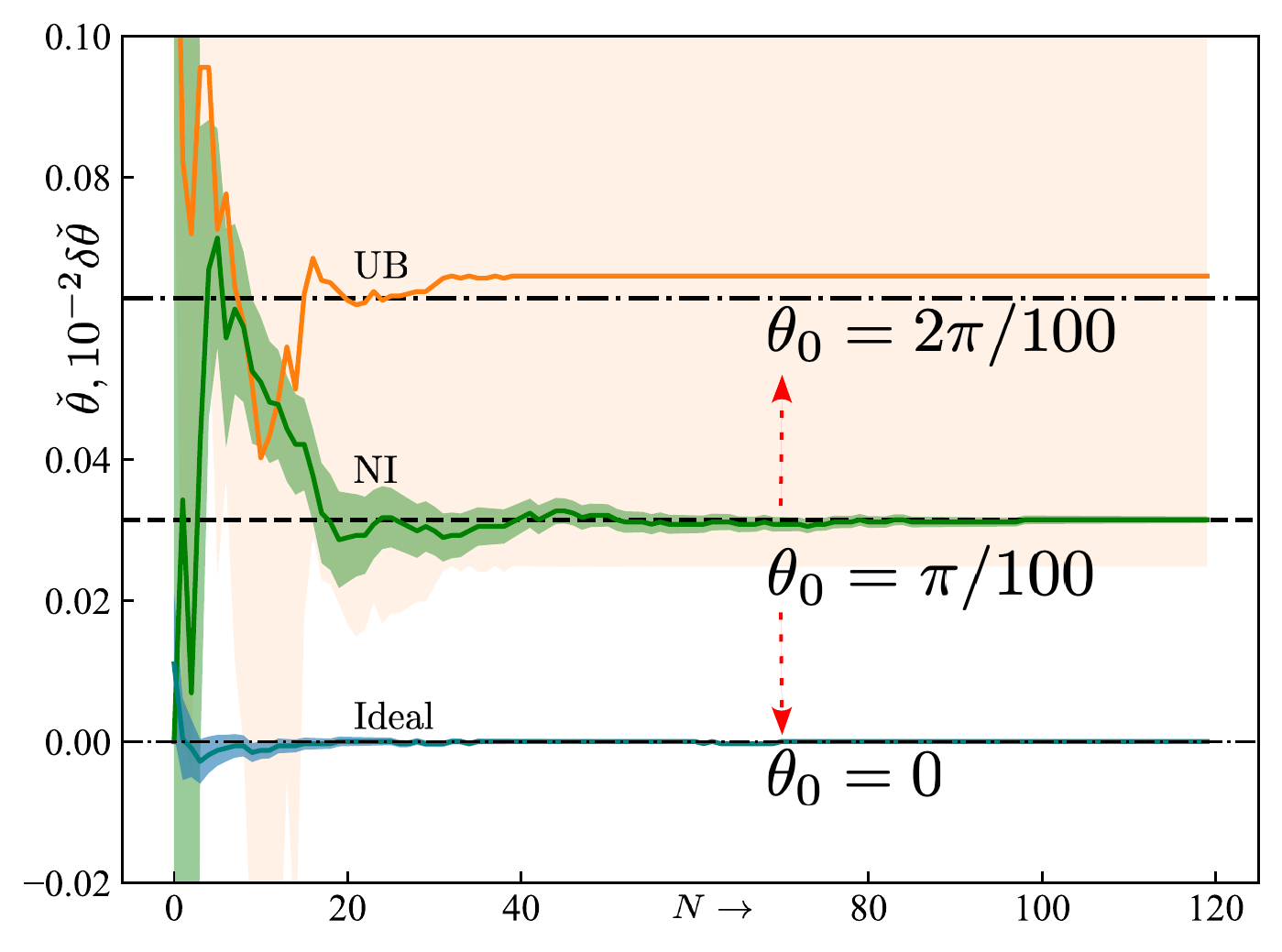}
    \caption{Maximum likelihood estimation (MLE) average $\check{\theta}$ and standard deviation $\delta\check{\theta}$ as number of measurements for the case of sequential metrology without resetting. Here the system and pointer state taken to be at $5\mathrm{GHz}$ are prepared in the initial thermal state corresponding to $\mathrm{T}_{\mathrm{S}} = 100 \mathrm{mK}$. The state of the system is evolved by the Hamiltonian $H = \theta \sigma_{x}$, where $\theta = \pi/100$ and measured along $z$ direction. MLE corresponding to (a) UB measurement (orange), (b) NI measurement (green) and (c) ideal measurement (teal) are presented alongside a dashed line that represents the true value $(\pi/100)$. Note that the trajectories corresponding to UB measurements have been displaced by $+\pi/100$ and likewise the trajectories of ideal measurement have been displaced by $-\pi/100$ as indicated on the figure above for clarity. Note that the UB estimate converges to the wrong value and its standard deviation, given by the shaded orange region, is large. On the other hand the ideal and the NI measurements converge to the correct mean with low standard deviations, and the ideal outperforms the NI measurement.} 
    \label{fig:3}
\end{figure}
The total $N_{s}\nu$ number of measurements performed result in $N$ sequences $\gamma^{(i)}=\{\nu^{(i)}_{\uparrow}/\nu,\nu^{(i)}_{\downarrow}/\nu\}$ which represent experimental frequencies. In the case of ideal measurements, since the system state and the pointer statistics are each unchanged at the end of the measurement, $N_{s}$ copies of the ideal measurement statistics $\gamma^{(i)}_{\text{id}}$ are received at the detectors. In contrast to this, the situation is different for non-ideal measurements. In the case of NI measurements, while the reduced system state remains the same at the end of each measurement, the pointer statistics $\gamma^{(i)}_{NI}$ are slightly different owing to the finite temperature discussed till now. This case is exactly reversed for UB measurements, where the reduced state of the system is modified at the end of each measurement whereas the pointer statistics $\gamma^{(i)}_{UB}$ remains unaltered. Hence each measurement considered above might acquire information about the unknown phase at a different rate. To study all three schemes fairly using the same analytical technique, we estimate the unknown phase by employing a maximum likelihood estimator (MLE) which considers all collected measurement statistics to infer the unknown parameter.

In Fig.~(\ref{fig:3}) we present the MLE estimate of parameter $\check{\theta}$ as a function of the number of measurements $N$. The log-likelihood function $l(\theta):=\sum_\alpha\gamma_\alpha\log(p_\alpha(\theta))$ is defined in terms of the the observed statistics $\gamma$ and the parametrized probability distribution $p(\theta)$ corresponding to the outcome index by $\alpha$.  This likelihood function can also be seen as the second term in a Kullback-Liebler divergence $\mathrm{KL}(\gamma\Vert p(\theta))$ and hence derives meaning as the minimization of a divergence between probability distributions. We note that for multiple measurements $\gamma^{(i)}$ resulting from potentially different parametrized probability distributions $p^{(i)}(\theta)$ defined as a function of the same unknown parameter $\theta$, the divergence compares the product distributions of the statistics and the parametrized distributions, leading to a log-likelihood function $l(\theta):=\sum_{i}\sum_\alpha\gamma^{(i)}_\alpha\log(p^{(i)}_\alpha(\theta))$. The estimate $\check{\theta}$ maximizes $l(\theta)$ and the standard deviation is computed via the curvature of the log-likelihood function as $(-\mathbb{E}[\nabla^2 l(\check{\theta})])^{-\frac{1}{2}}$.
In Fig~(\ref{fig:3}), we estimated the parameter whose true value was $\theta = \pi/100$ using MLE method for the sequential metrological scheme discussed above for ideal, NI and UB measurements.  In the case of ideal measurements, we observe that $\check{\theta}$ saturates to $\theta_{true}$ after few measurements. Furthermore, the standard deviation in the estimate also converges to attainable sensitivity after the few iterations.  This serves as a benchmark for the performance of the other two non-ideal measurement schemes. On the other hand, the estimated value from UB measurement saturates to a wrong value.  We understand this as the result of the reduced system state approaching maximally mixed state after repeated measurements. For instance, the purity of the post-measurement states after $N_s=120$ measurements is $\approx0.8$ for NI whereas it is $=0.5$ for UB up to numerical accuracy. Hence, its ability to acquire the correct information about the parameter deteriorates after each measurement.  The standard deviation of the estimator is also very large for the same reason.  Finally for NI measurement, the system state is unchanged after each measurement, so the estimated value converges to  $\theta_{true}$.  The standard deviation of the estimator also approaches better sensitivity after few measurements, but less than ideal measurement case. In summary, NI measurements favor schemes such as repeated metrology without resetting. This is because preserving the statistics of the states after each measurement is more important for such schemes. 
\section{Discussion and Conclusion}
An ideal measurement is thermodynamically inconsistent as it requires infinite resources. The inclusion of thermodynamics gives rise to two different realistic measurements namely UB and NI measurements.
We showed that for UB measurement, WVA shows better amplification by consuming fewer resources, which is result of pointer replicating system statistics.
Due to this fact, the amplification for UB measurement depends only on the system temperature and depends on both system and pointer temperature for NI measurement. 
We also considered a single-shot sequential metrological scheme in which the NI measurement shows the advantage over UB measurement.  This is because we considered a scheme without resetting the probe state and we see that preserving the purity of the reduced state of the system is important to the task of acquiring information about the phase over successive runs.

We have discussed the strict constraints that thermodynamically consistent measurements place on quantum experiments, even though such experiments have been performed before \cite{dixon, Knee2016, WVA_expt_photon_recycling} and requires some discussion. It should be noted that the imperfect correlation between system and pointer is a consequence of a full rank finite dimensional pointer at non-zero temperature. This analysis can hence be circumvented by the use of an infinite dimension pointer \cite{guryanova2020ideal} such as the harmonic oscillator or by cooling to a low enough temperature as noted in Fig.~(\ref{fig:1}). In reality, using infinite dimensional pointers such as mechanical or light modes is not guarenteed and further care has to be taken.  

Hence our analysis suggests that realistic metrological tasks with thermodynamically consistent measurements can be classified into two categories based on whether the pointer statistics is important or the reduced states of the system. Finally we note that optimizing over other figures of merits might posit schemes which use UB and NI measurements together to acheive optimal performance. Our results hence contribute to the general task of designing realistic quantum metrology tasks with thermodynamically consistent resources and can guide such experimental design in the future.

\begin{acknowledgments}
\paragraph*{Acknowledgments.---} S.V. acknowledges support from Government of India DST-QUEST grant number DST/ICPS/QuST/Theme-4/2019 and discussions with R.Vijay and F.Binder. 
\end{acknowledgments}
\bibliographystyle{apsrev4-1}

%

\newpage
\appendix

\section{Proof of Non-ideal Weak Value}
In this section, we evaluate the kicked state for non-ideal measurements.  Consider the initial state of the system is prepered in thermal state state 

\begin{equation}
\rho_{s} = \Tilde{p} \ket{\psi_{i}}\bra{\psi_{i}} + (1-\Tilde{p})\tau, \label{ni1}
\end{equation}
where $(\Tilde{p}/(1-\Tilde{p}) = \exp(-\beta E_{s}))$.  Now the initial state can be re-written as

\begin{equation}
    \rho_{s} = q \ket{\psi_{i}}\bra{\psi_{i}} + \Bar{q} \ket{\psi_{i}^{\perp}}\bra{\psi_{i}^{\perp}}. \label{ni2}
\end{equation}

the combined system and meter state is

\begin{eqnarray}
    \rho_{sm} & = & \Bar{q} e^{-igA\otimes B} (\ket{\psi_{i}}\bra{\psi_{i}}\otimes \rho_{m}) e^{igA\otimes B} \nonumber \\
    & & +\Bar{q} e^{-igA\otimes B} (\ket{\psi_{i}^{\perp}}\bra{\psi_{i}^{\perp}}\otimes \rho_{m}) e^{igA\otimes B}. \label{ni3}
\end{eqnarray}

Using the Gram-Schmit orthogonalization procedure, writing the initial state in terms of final states, we get
\begin{eqnarray}
\ket{\psi_{i}} & = & a \ket{\psi_{f}} + b \ket{\psi_{f}^{\perp}}, \nonumber\\
\ket{\psi_{i}^{\perp}} & = & b^{*}\ket{\psi_{f}} - a^{*} \ket{\psi_{f}^{\perp}}, \label{ni4}
\end{eqnarray}

where $a = \braket{\psi_{f} \vert \psi_{i}}$.  Finally 
\begin{eqnarray}
\ket{\psi_{i}}\bra{\psi_{i}} & = &  \vert a\vert^2 \ket{\psi_{f}}\bra{\psi_{f}} + ab^{*} \ket{\psi_{f}}\bra{\psi_{f}^{\perp}} \nonumber\\
& & a^{*}b \ket{\psi_{f}^{\perp}}\bra{\psi_{f}} + \vert b \vert^2 \ket{\psi_{f}^{\perp}}\bra{\psi_{f}^{\perp}} \label{ni5}
\end{eqnarray}

and

\begin{eqnarray}
\ket{\psi_{i}^{\perp}}\bra{\psi_{i}^{\perp}} & = &  \vert b\vert^2 \ket{\psi_{f}}\bra{\psi_{f}} - ab^{*} \ket{\psi_{f}}\bra{\psi_{f}^{\perp}} \nonumber\\
& &- a^{*}b \ket{\psi_{f}^{\perp}}\bra{\psi_{f}} + \vert a \vert^2 \ket{\psi_{f}^{\perp}}\bra{\psi_{f}^{\perp}}. \label{ni6}
\end{eqnarray}

The total system-meter state can be rewritten in terms of final state and state orthogonal to the final state

\begin{eqnarray}
\rho_{sm} & = & {q} [(\vert a \vert^2 \ket{\psi_{f}}\bra{\psi_{f}}+\vert b \vert^2 \ket{\psi_{f}^{\perp}}\bra{\psi_{f}^{\perp}})\otimes \rho_{m} \nonumber\\
& & - ig (ba^{*}\ket{\psi_{f}}\bra{\psi_{f}}-ab^{*}\ket{\psi_{f}^{\perp}}\bra{\psi_{f}^{\perp}})\otimes B \rho_{m} \nonumber\\ 
& & + ig (a^{*}b \ket{\psi_{f}^{\perp}}\bra{\psi_{f}^{\perp}}-ab^{*}\ket{\psi_{f}}\bra{\psi_{f}}) \otimes \rho_{m}B] \nonumber\\
& &  \Bar{q} [(\vert b \vert^2 \ket{\psi_{f}}\bra{\psi_{f}}+\vert a \vert^2 \ket{\psi_{f}^{\perp}}\bra{\psi_{f}^{\perp}})\otimes \rho_{m} \nonumber\\
& & + ig (ba^{*}\ket{\psi_{f}}\bra{\psi_{f}}-ab^{*}\ket{\psi_{f}^{\perp}}\bra{\psi_{f}^{\perp}})\otimes B \rho_{m} \nonumber\\ 
& &- ig (ab^{*} \ket{\psi_{f}^{\perp}}\bra{\psi_{f}^{\perp}}-a^{*}b\ket{\psi_{f}}\bra{\psi_{f}}) \otimes \rho_{m}B] \nonumber\\
& &+\text{off-diagonal terms}. \label{ni7}
\end{eqnarray}
Here and below off-diagonal terms do not come in calculations because they are eliminated after measurement. Now we take a thermal pointer state and correlate it with system only on system-meter state by either unbiased method or non-invasive method.
The correlation matrix for unbiased is 

\begin{equation}
    U_{UB} = \begin{bmatrix}
               1 & 0 & 0 & 0\\
               0 & 0 & 0 & 1\\
               0 & 1 & 0 & 0\\
               0 & 0 & 1 & 0\\
               \end{bmatrix}
\end{equation}

and correlation matrix for the non-invasive method is

\begin{equation}
    U_{NI} = \begin{bmatrix}
               1 & 0 & 0 & 0\\
               0 & 1 & 0 & 0\\
               0 & 0 & 0 & 1\\
               0 & 0 & 1 & 0\\
               \end{bmatrix}.
\end{equation}

Then correlating the joint system and meter with the evolution gives
\begin{equation}
  \rho_{psm} =  U_{UB/NI} (\rho_{p}\otimes\rho_{sm}) U_{UB/NI}^{\dagger}. \label{ni8} 
\end{equation}

For unbiased, the correlated state is

\begin{eqnarray}
  \Psi^{ps}(\rho_{psm}) & = & p P_{s} \Bar{q} \ket{\psi_{f}}\bra{\psi_{f}} \otimes \ket{\psi_{f}}\bra{\psi_{f}}\otimes \eta_{1} \nonumber\\ 
  & & + p\Bar{P}_{s}\Bar{q} \ket{\psi_{f}}\bra{\psi_{f}} \otimes \ket{\psi_{f}}\bra{\psi_{f}} \otimes \eta_{2} \nonumber\\
  & & + \Bar{p}P_{s} \Bar{q} \ket{\psi_{f}}\bra{\psi_{f}}\otimes \ket{\psi_{f}^{\perp}}\bra{\psi_{f}^{\perp}} \otimes \eta_{1} \nonumber\\
  & & + \Bar{p}\Bar{P}_{s}\Bar{q}\ket{\psi_{f}}\bra{\psi_{f}}\otimes \ket{\psi_{f}^{\perp}}\bra{\psi_{f}^{\perp}}\otimes \eta_{2} \nonumber\\  & &+\text{off-diagonal terms}. \label{ni10}
\end{eqnarray}

For NI measurement, the correlated state is

\begin{eqnarray}
\Psi^{ps}(\rho_{psm}) & = & p P_{s} \Bar{q} \ket{\psi_{f}}\bra{\psi_{f}} \otimes \ket{\psi_{f}}\bra{\psi_{f}}\otimes \eta_{1} \nonumber\\
& & + p\Bar{P}_{s}\Bar{q} \ket{\psi_{f}}\bra{\psi_{f}} \otimes \ket{\psi_{f}}\bra{\psi_{f}} \otimes \eta_{2} \nonumber\\
& & + \Bar{p}\Bar{P}_{s} \Bar{q} \ket{\psi_{f}}\bra{\psi_{f}}\otimes \ket{\psi_{f}^{\perp}}\bra{\psi_{f}^{\perp}} \otimes \eta_{1} \nonumber\\
& & + \Bar{p}P_{s}\Bar{q}\ket{\psi_{f}}\bra{\psi_{f}}\otimes \ket{\psi_{f}^{\perp}}\bra{\psi_{f}^{\perp}}\otimes \eta_{2} \nonumber \\
& &+ \text{off-diagonal terms}, \label{ni11}
\end{eqnarray}

where $\eta_{1} = \exp({-ig\mathcal{A}_{\mathrm{w}}B})\rho_{m}\exp({ig\mathcal{A}_{\mathrm{w}}B})$ and $\eta_{1} = \exp({-i({g}/{\mathcal{A}_{\mathrm{w}}})B})\rho_{m}\exp({+i({g}/{\mathcal{A}_{\mathrm{w}}})B})$, such that $({g}/{\mathcal{A}_{\mathrm{w}}}) \ll g\mathcal{A}_{\mathrm{w}} \ll 1$.

Finally measurement in pointer onto $\ket{\psi_{f}}$ state, the post-measurement meter state for UB measurement is

\begin{equation}
    \rho_{m}^{PS(UB)} = Tr_{p}(\Psi_{pm}^{unb}) = (qP_{s}) \eta_{1} + (\Bar{P}_{s}\Bar{q}) \eta_{2} \label{ni12}
\end{equation}

and for NI measurement is
\begin{eqnarray}
    \rho_{m}^{PS(NI)}  =  Tr_{p}(\Psi_{pm}^{ni}) & = & (pqP_{s} \eta_{1}+\Bar{p}\Bar{q}P_{s} \Tilde{\eta}_{1}) \nonumber\\ 
    & &+ (p\Bar{P}_{s}\Bar{q}+\Bar{p}q\Bar{P}_{s}) \eta_{2}. \label{ni12}
\end{eqnarray}

where $\Tilde{\eta}_{1} = \exp({ig\mathcal{A}_{\mathrm{w}}B})\rho_{m}\exp({-ig\mathcal{A}_{\mathrm{w}}B}) $ which is a Gaussian centered at $x = -g\mathcal{A}_{\mathrm{w}}$.

We see that the amplification will be better for UB measurement compared to NI measurement because the accurate amplification will be a weighted average of the spread in the meter space.  For UB measurement, the weighted average lies between $x = 0$ and $x=g\mathcal{A}_{\mathrm{w}}$.  For NI measurement, the weighted average is nearly close to $x=0$.  If we imagine the meter state is a Gaussian function in the position-basis, we see two Gaussians, one centered at $x=0$ due to $\eta_{2}$, and another Gaussian centered at $x=g\mathcal{A}_{\mathrm{w}}$ due to $\eta_{1}$.

Now we calculate the Fisher information for a non-ideal measurement and compare it to Fisher information obtained from an ideal measurement.  The post-selected state for a non-ideal measurement state can be re-written as

\begin{equation}
    \rho(g) = P_{M} (e^{-ig\mathcal{A}_{\mathrm{w}}B}\rho_{m}e^{ig\mathcal{A}_{\mathrm{w}}})+\gamma \rho_{m}.  \label{mfi1}
\end{equation}

Now diagonalizing Eq.~\eqref{mfi1}, so that we can calculate Bures distance easily

\begin{eqnarray}
    \rho(g) & = & \frac{P_{M}}{2} [\vert g\mathcal{A}_{\mathrm{w}}\vert^2 \frac{2\gamma}{\gamma+1}\ket{\psi_{1}}\bra{\psi_{1}}\nonumber\\ 
    & &+(\gamma+1)\frac{\vert g \mathcal{A}_{\mathrm{w}}\vert^2}{\gamma+1}\ket{\psi_{2}}\bra{\psi_{2}}].
\end{eqnarray}

Now the Bures distance can be calculated from the fidelity between between $\rho_{g}$ and $\rho_{g+dg}$  

\begin{eqnarray}
    &F(\rho_{g}, \rho_{g+dg})   =  \vert g \mathcal{A}_{\mathrm{w}} \vert \vert (g+dg) \mathcal{A}_{\mathrm{w}}\vert \frac{\gamma}{\gamma+1} \nonumber\\
     & + (\gamma+1) \sqrt{(1+\vert g\mathcal{A}_{\mathrm{w}}/(\gamma+1)\vert^2)\nonumber}\\
     &\sqrt{(1+\vert (g+dg)\mathcal{A}_{\mathrm{w}}/\gamma+1\vert^2)}.
\end{eqnarray}

Computing the negative second derivative of $g$ of Bures distance gives us Fisher information.

\begin{equation}
    \mathcal{I}(g) = -\partial_{g}^2 [2(1-F(\rho_{g}, \rho_{g+dg}))].
\end{equation}

\begin{equation}
    \mathcal{I}_{TH}(g) = 4P_{M} \vert\mathcal{A}_{\mathrm{w}}^{'}\vert^2 (1-\vert g \mathcal{A}_{\mathrm{w}}^{'}\vert^2).
\end{equation}

where the weak value is
\begin{equation}
\mathcal{A}'_{\mathrm{w}}  =  \frac{\mathcal{A}_{\mathrm{w}}}{1+\gamma_{M}} .
\end{equation}

and $\delta_{M}$ is 
\begin{equation}
\delta_{M} = \begin{cases} 
\frac{\Bar{q}\Bar{P}_{s}}{qP_{s}} &  \text{for unbiased,} \\

\frac{p\Bar{P}_{s}\Bar{q}+\Bar{p}q\Bar{P}_{s}}{pqP_{s}+\Bar{p}\Bar{q}P_{s}} & \text{for noninvasive}.
\end{cases}
\end{equation}

and $P_{M}$ is 
\begin{equation}
P_{M} = \begin{cases} 
qP_{s} &  \text{for unbiased,} \\
pqP_{s}+\Bar{p}\Bar{q}P_{s} & \text{for noninvasive}.
\end{cases}
\end{equation}

These are Eq.(\ref{qf1})-Eq.(\ref{eq14}) in the main text.  The amplification is bad for both non-invasive and unbiased measurements compared to an ideal measurement.  If you compare the NI method with the UB, the Fisher information obtained from UB measurement is better than the other method due to the amplification depending only on the system purity, not on the pointer purity. Because in UB measurement, the pointer replicates the system statistics which is more important in weak value amplification procedure.  THe NI procedure only preserves the states of the system and the system statistics are changed, hence we got better amplification through an UB measurement procedure compared to NI measurement.

\end{document}